\documentclass[aps,prl,twocolumn,notitlepage]{revtex4-1}

\usepackage{amstext,amsmath,amssymb,amsfonts,bbm}
\usepackage[latin1]{inputenc}
\usepackage{epsfig}
\usepackage{hyperref}
\usepackage{amsthm}
\usepackage{subfigure}
\usepackage{color}
\usepackage{multirow}

\newcommand{\bea}{\begin{eqnarray}}
\newcommand{\eea}{\end{eqnarray}}
\newcommand{\be}{\begin{equation}}
\newcommand{\ee}{\end{equation}}
\newcommand{\cG}{{\cal G}}

\newcommand{\cT}{{\cal T}}

\newcommand{\cF}{{\cal F}}
\newcommand{\cB}{{\cal B}}
\newcommand{\cL}{{\cal L}}
\newcommand{\cN}{{\cal N}}
\newcommand{\cJ}{{\cal J}}
\newcommand{\cH}{{\cal H}}
\newcommand{\cS}{{\cal S}}

\newtheorem{lemma}{Lemma}

\newtheorem{definition}{Definition}
\newtheorem{theorem}{Theorem}

\begin{document}

\title{The $1/N$ expansion of colored tensor models in arbitrary dimension}

\author{Razvan Gurau${}^a$}\email{rgurau@perimeterinstitute.ca}
\author{Vincent Rivasseau$^{b}$}\email{rivass@th.u-psud.fr}

\affiliation{${}^a$Perimeter Institute for Theoretical Physics, 31 Caroline
St, Waterloo, ON, Canada \\
${}^b$Laboratoire de Physique Th\'eorique, Universit\'e Paris XI,
Orsay, France.
}

\begin{abstract}
\noindent In this paper we extend the $1/N$ expansion introduced in \cite{colorN} to 
group field theories in arbitrary dimension and prove that only graphs 
corresponding to spheres $S^D$ contribute to the leading order 
in the large $N$ limit.
\end{abstract}

\maketitle

\section{Introduction}

The $1/N$ expansion, together with Wilson's-Fisher $\epsilon=4-D$ expansion around mean field theory
and the $1/D$ expansion of condensed matter around dynamical mean field theory, stands out as the main
alternative to the ordinary coupling constant expansion or to numerical simulations
in quantum field theory and statistical mechanics. The (large) parameter $N$ is a range of integer values
for fields indices encoding some tensorial structure.
Until now the $1/N$ expansion was understood solely
for vector or matrix models but not for higher rank tensor fields.

The leading terms in this expansion for the free energy
are linear chains (rings) made of a simple bubble motive 
for vector models \cite{chain} and planar graphs for matrix models \cite{planar, Brez}. The corresponding Feynman
graphs obviously pave respectively the circle $S^1$ and the 2 dimensional sphere $S^2$. It is therefore
tempting to conjecture that higher rank tensor fields with $D$ indices should lead to models
which admit also a $1/N$ expansion in which
dominant graphs pave the $D$ dimensional sphere $S^D$.
However no examples of such a higher rank $1/N$ expansion had been found until recently.

In \cite{colorN} the first expansion of this type was found for 
the {\it colored} \cite{color} Boulatov model \cite{Boul} (which is a group field theory in $D=3$).
The color condition seems a key ingredient for this $1/N$ expansion
to hold, or at least to be tractable. Its graphs are duals to
manifolds with only point-like singularities \cite{lost}.
Remarkably, the colors are also necessary to implement the diffeomorphism symmetry in GFT \cite{Baratin:2011tg}.
The main reason for the interest in group field theory lies in its potential for quantizing gravity \cite{laurentgft,quantugeom2}.
Just as matrix models are related to 2-D gravity \cite{mm,Di Francesco:1993nw}, it is expected that $D$-rank
group field models are related to $D$-dimensional gravity \cite{ambj3dqg,mmgravity}.
The physically interesting case is then $D=4$, hence the $1/N$ expansion 
for rank-four tensors is even more important than the one for rank-three tensors.

Ideally we would like quantum gravity to be based on a simple quantum
field theory, since this is the case for all other known forces of physics. But we 
would like also to understand
why space-time, which is expected to be wild and foamy at the Planck scale
or beyond, is so smooth and manifold-like at large distances. Some mechanism should
flatten out the virtual loops and handles of quantum space-time and favor the 
trivial $S^4$ topology, since this is the long-distance classical world that we actually observe.

In this letter we extend the method of \cite{colorN} to prove that the {\it colored}
group field theory in $D$ dimensions
indeed admits a $1/N$ expansion dominated by graphs which pave the $S^D$ manifold.
This includes the physically interesting case of $D=4$, namely 
the colored topological Ooguri model. Realistic models for quantum gravity should include dynamical degrees of freedom,
as is attempted eg. in the EPR-FK models \cite{newmo2,newmo3,newmo4,newmo5}, and we hope some more complicated form of the $1/N$ 
expansion can be extended also to the colored version of these models. 

We would like to stress that, to our knowledge, the results in this paper give the first analytic 
example of such a dominance of $S^4$ topologies in a quantum field theory model. 
Dominance of large and smooth structures is indeed not easy to obtain in
sums over random space-times: most naive models
tend to develop crumpled or polymer-like phases. To our knowledge the only other
approach in which simulations of random space-times lead to large and smooth structures is 
causal dynamical triangulations \cite{AJL}, but they are based exclusively on numerical 
and not analytic results.

Our result is the following.
We consider the $D$-dimensional GFT theory introduced in \cite{color}.
The corresponding $D+1$ colors are noted as 0,1 ... $D$, and the vertices are noted sloppily
as $\lambda \phi^{D+1}$ and $\bar \lambda (\bar \phi)^{D+1}$. 
Introducing a cutoff on the group representations
provides the index $N$ (i.e. we suppress representations of spin higher than $N$). 
The corresponding regularized $\delta$ function on the group
diverges at the origin when $N \to \infty$ as $\delta^N (e)$. 
We prove below that the free energy $F = \log Z$ of the theory obeys
\bea \label{main}
 F = \delta^N(e)^{D-1} C(\lambda,\bar\lambda) + O [ \delta^N(e)^{D-1-\frac{2(D-2)}{D!}} ] \; ,
\eea
where all graphs contributing to $C(\lambda,\bar\lambda) $ have the topology of spheres $S^D$.

We {\it do not} prove that the subleading terms in \eqref{main} are also indexed by topologies. We explain why 
in $D=3$ it has been possible to go further and prove that the entire series is index by topologies (encoded in 
the {\it core graphs} of \cite{colorN}). We expect an analog of this to be also true in arbitrary $D$, but
establishing this is beyond the scope of this paper.

\section{Graphs, Jackets and Bubbles}

\begin{figure}[htb]
\begin{center}
 \includegraphics[width=1.5cm]{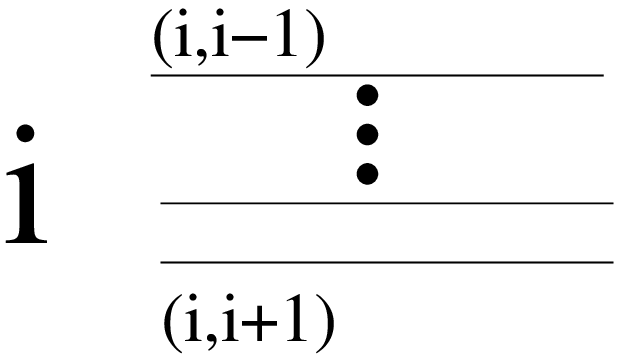} \hspace{0.8cm}
 \includegraphics[width=1.5cm]{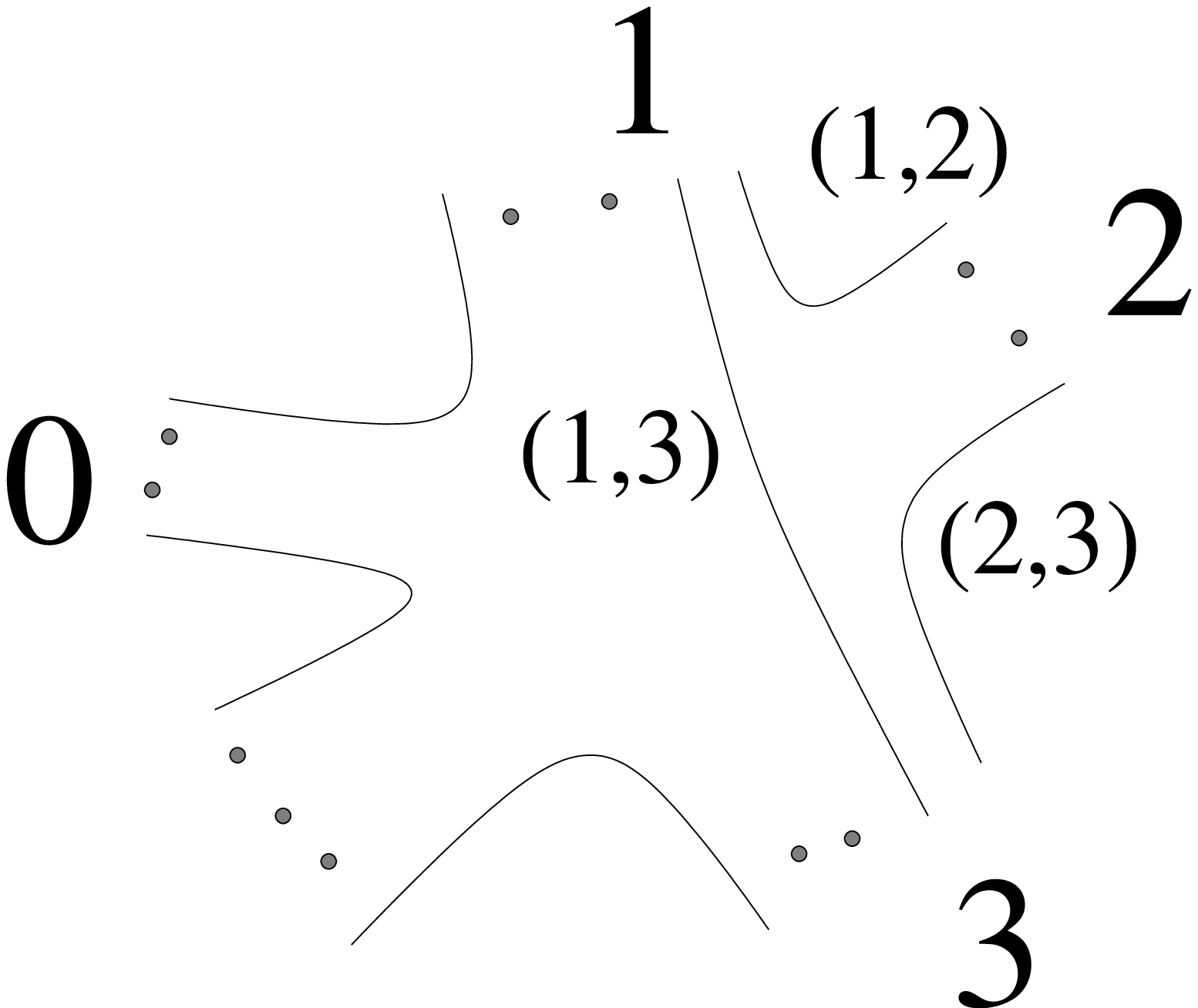} 
\caption{Line and vertex of the Colored GFT graphs.}
\label{fig:propvert}
\end{center}
\end{figure}

Let $G$ be some compact multiplicative Lie group, and denote $h$ its elements,
$e$ its unit, and $\int dh$ the integral with respect to the Haar measure.
Let $\bar \psi^i,\psi^i$,  $i=0,1,\dots, D$ be $D+1$ couples of complex 
scalar (or Grassmann) fields over $D$ copies of $G$, $\psi^i:G^{\times D} \rightarrow \mathbb{C}$. 
Denote $ \psi(h_0,\dots, h_{D-1}):=\psi_{h_0,\dots, h_{D-1}}$.
The partition function of the $D$ dimensional colored GFT \cite{color,lost,colorN} is defined by the path integral 
\bea
 e^{-F} = Z(\lambda,\bar\lambda) = \int \prod_{i=0}^D d\mu_P(\psi^i,\bar\psi^i) \; e^{-S^{int}-\bar S^{int}}
\; ,
\eea
with normalized Gaussian measure of covariance $P$ and interaction $S^{int}$
\bea\label{eq:action}
&& P_{h_{0}\dots h_{D-1} ; h_{0}'\dots h_{D-1}'} 
 = \int dh \;  \prod_{i=0}^{D-1} \delta^{N}\bigl( h_{i} h (h_{i}')^{-1} \bigr) \; ,\crcr
&& S^{int} = \frac{\lambda}{ \sqrt{ \delta^N(e)^{\frac{(D-2)(D-1)}{2}} } } 
\int \prod_{i<j} dh_{ij} \crcr
&&\qquad \prod_{i=0}^{D+1} \psi^i_{h^{}_{ii-1} \dots h^{}_{i0} h^{}_{in} \dots h_{ii+1}}  \; , 
\eea
where $h_{ij} = h_{ji}$. The index $i\in \{0,\dots,D\}$ of each field is a {\it color} index. 
The Feynman graphs of the D-dimensional colored GFT (called (D+1)- colored graphs), 
are made of oriented colored lines with $D$ parallel threads and oriented vertices (dual to $D$ simplices) of
coordination $D+1$. For every vertex, the thread $(i,j)$ connects the halflines of colors $i$ and $j$ 
(see figure \ref{fig:propvert}). The $3$-colored graphs are the familiar ribbon graphs of matrix models.

Colored graphs \cite{color,lost} are dual to normal pseudo manifolds. 
The free energy $F$ of the colored GFT writes as a sum over connected vacuum 
colored graphs $\cG$. We denote 
$\cN_{\cG},\; |\cN_{\cG}|=2p$, $\cL_{\cG}$, $\cF_{\cG}$, the sets of vertices, lines and  faces 
(i.e. closed threads) of $\cG$. The amplitude of $\cG$, $A^{\cG}$ is
\cite{color,colorN,sefu3}
\bea  \label{eq:ampli}
\frac {  (\lambda\bar\lambda)^p}
{ [\delta^N(e)]^{p\frac{(D-2)(D-1)}{2} } }  
\int \prod_{\ell\in \cL_{\cG}} dh_{\ell} 
\prod_{f\in \cF_{\cG}} \delta^N_{f}(\prod_{\ell\in f }^{\rightarrow} h_{\ell}^{\sigma^{ \ell | f}} )
\; ,
\eea
where $\sigma^{\ell|f}=1$ (resp. $-1$) if the orientations of 
$\ell$ and $f$ coincide (resp. are opposite) and $\sigma^{\ell|f}=0$ if $\ell$ does not 
belong to the face $f$.

\begin{figure}[htb]
\begin{center}
 \includegraphics[width=1.5cm]{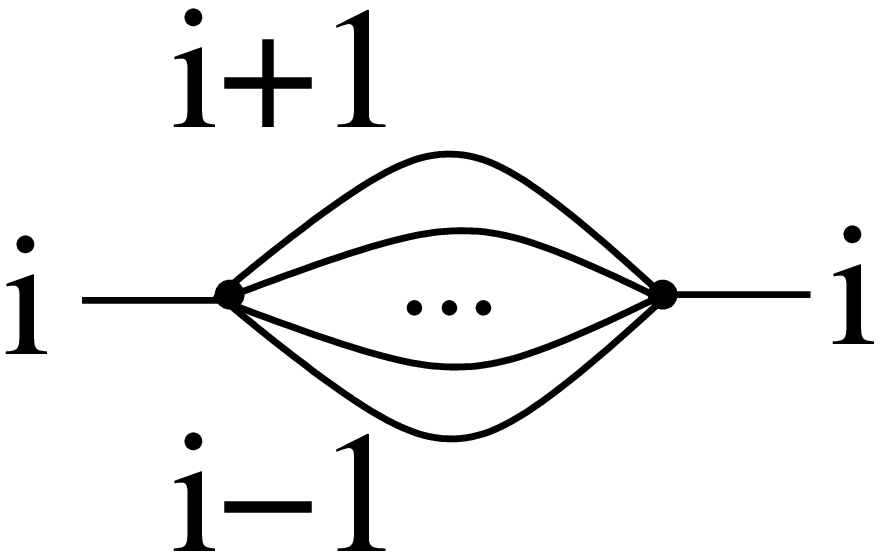}  
\caption{The two point graph $\cG_1$.}
\label{fig:melon}
\end{center}
\end{figure}

Consider the two point colored graph with two vertices connected by $D$ lines (denoted $\cG_1$)
represented in figure \ref{fig:melon}. For any graph $\cG$, one can consider
the family obtained by inserting $\cG_1$ an arbitrary number of times on 
any line of $\cG$ . The scaling of the coupling constants in eq. \eqref{eq:action} is the only scaling 
which ensures that this family has uniform degree of divergence, hence it is the only scaling 
under which a $1/N$ expansion makes sense.

To a colored graph $\cG$ one associates two categories of subgraphs: its bubbles \cite{color, lost} and 
its jackets \cite{colorN, sefu3}.
The $0$-bubbles of $\cG$ are its vertices and the $1$-bubbles are its lines.
For $p\ge 2$, the {\it $p$-bubbles} with colors $\{i_1,\dots,i_p\}$ of $\cG$
are the connected components (labeled $\rho$) obtained from $\cG$
by deleting the lines and faces containing at least one of the colors $\{0,\dots,D\} \setminus \{i_1,\dots,i_p\}$. 
We denoted the p-bubbles $\cB^{i_1\dots i_p}_{(\rho)}$.  
For $p\ge 2$ each $p$-bubble is a $p$-colored graph, and the $2$-bubbles are
the faces of $\cG$. 

\begin{figure}[htb]
\begin{center}
 \includegraphics[width=3cm]{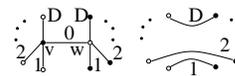}  
\caption{Contraction of a 1-Dipole.}
\label{fig:1canc}
\end{center}
\end{figure}

We denote $\widehat{i}= \{0,\dots,D\} \setminus \{i\}$. 
Consider a line (say of color $0$) $l^0$ with end vertices $v$ and $w$ 
in a graph $\cG$. Each of the vertices $v$ and $w$ belongs to some $D$-bubble of colors 
$\widehat{0}$, $\cB^{\widehat{0}}_{(\alpha)} $ and $\cB^{\widehat{0} }_{ (\beta) }$. If the
two bubbles are {\it different} and at least one of them is dual to 
a {\it sphere} $S^{D-1}$, then $l^0$ is a {\bf 1-Dipole} 
\cite{colorN,FG,Lins}. A 1-Dipole can be contracted, that is the line $l^0$ 
together with the vertices $v$ and $w$ can be deleted from the graph 
and the remaining lines reconnected {\it respecting the coloring} 
(see figure  \ref{fig:1canc}). The fundamental result we will use in the 
sequel \cite{FG} is that the two pseudo manifolds dual to
$\cG$ and $\cG'$ are homeomorphic if $\cG$ and $\cG'$ are related by a 
1-Dipole contraction. We call two such graphs ``equivalent'', $\cG \sim \cG'$.

\begin{definition}
A colored {\bf jacket} $\cJ$ of $\cG$ is the ribbon graph made by the faces of colors 
$(\tau^k(0),\tau^{k+1}(0))$, for some cycle $\tau$ of $D+1$ elements,
modulo the orientation of the cycle. We denote $J=\{ (\tau^k(0), \tau^{k+1}(0)) \;  | k=0,\dots, D \}$
the set of faces of $\cJ$.
\end{definition}

For $D=2$ the (unique) jacket of a colored graph is the graph itself.
The reader can check that $\cJ$ and $\cG$ have the same connectivity, the number of jackets
of a $D+1$ colored graph is $ \frac{1}{2}D!$ and the number of jackets containing a given pair 
is $(D-1)!$. In $D=4$ there are 5 colors, 
10 pairs of colors and 12 jackets (corresp. to the cycles $(01234)$, $(01243)$, etc.). 
The ribbon lines of the jacket $\cJ$ separate two faces, $(\tau^{-1}(j),j)$ and $(j,\tau(j))$ 
and inherit the color $j$ of the line in $\cG$. As the $p$-bubbles 
are $p$-colored graphs they also possess jackets which can be obtained from 
the jackets of $\cG$. 

For a jacket $\cJ$, $J^{\widehat{i}} = J \setminus \{(\tau^{-1}(i), i), (i,\tau(i))\} \cup \{ (\tau^{-1}(i), \tau(i)) \} $ 
is a cycle over the $D$ elements $\widehat{i}$. The ribbon subgraph of $\cG$ made of faces in
$J^{\widehat{i} }$ is the union of several connected components, $\cJ^{\widehat{i}}_{(\rho)}$. Each
$\cJ^{\widehat{i}}_{(\rho)}$ is a jacket of the $D$-bubble 
$\cB^{ \widehat{i} }_{ (\rho) }$. Conversely, every jacket of $\cB^{ \widehat{i} }_{ (\rho) }$ is obtained 
from exactly $D$ jackets of $\cG$ corresponding to inserting the color $i$ anywhere along the cycle of $D$ elements.

\begin{lemma}\label{lem:planar}
 If a jacket $\cJ$ is planar then, for all $i$ and $\rho$, the jacket graphs $\cJ^{\widehat{i}}_{(\rho)}$ 
of the $D$-bubbles  are planar.
\end{lemma} 

\noindent{\bf Proof:} We delete one by one all lines of color $i$. 
We denote $\cJ \setminus l^i$ the graph obtained from $\cJ$ by deleting the line $l^i$.
As long as the two faces of $\cJ\setminus l^i_1 \dots \setminus l^i_{p-1}$ 
touching $l^i_p$ are different, the number of faces decreases by $1$ when deleting $l^i_p$, 
hence the Euler character $\chi$ of $\cJ \setminus l^i_1 \dots \setminus l^i_{p-1}$ 
equals the Euler character $\chi'$ of $\cJ \setminus l^i_1 \dots \setminus l^i_{p-1} \setminus l^i_p$ .

Suppose that we reach a line $l_{p}^i$ such that the two sides of the ribbon belong to the same face
of $\cJ \setminus  l^i_1 \dots \setminus l^i_{p-1}$. Erasing $l^i_p$ increases the number of faces 
by one, hence $\chi' = \chi +2$. As the Euler character factors over connected components, we conclude 
that erasing $l^i_p$ necessarily divides some (planar) connected component of the 
$\cJ \setminus  l^i_1 \dots \setminus l^i_{p-1}$ into two planar connected components of 
$\cJ \setminus l^i_1 \dots \setminus l^i_{p-1} \setminus l^i_p$. Hence all
$\cJ^{\widehat{i} }_{(\rho)}$ are planar. 

\qed

\section{The $1/N$ expansion}

All the group elements of the lines in a tree $\cT\in \cG$ can be eliminated from the amplitude 
\eqref{eq:ampli} by a tree change of variables \cite{FreiGurOriti}.
Consider a jacket $\cH$ of $\cG$. The lines of $\cH$ admit (many) partitions in three disjoint 
sets: a tree $\cT$ in $\cH$, a tree $\tilde \cT$ in the dual graph $\tilde \cH $, 
and a set $\cL \setminus \cT \setminus \tilde \cT$, of ``genus'' lines \cite{param}. 
Denoting $l(f,\tilde \cT)$ the line in the dual tree $\tilde \cT$ 
touching the face $f$ and going towards the root face $r$, the contribution of the faces of $\cH$ 
to the amplitude \eqref{eq:ampli} can be cast into the form \cite{colorN}
\bea \label{eq:routing}
&& \prod_{f\in \cH} \delta^N_{f}( \prod_{\ell}^{\rightarrow} h_{\ell}^{\sigma^{\ell | f}}) =
 \delta^N_{r} (\prod_{\ell \notin \tilde \cT }^{\rightarrow} h_{\ell}^{\sigma^{\ell | \cup_{f\in \cH} f} } )
\crcr
&& \qquad \times \prod_{f\in \cH, f\neq r} 
\delta^N_{f} \Big{(} h_{l(f,\tilde \cT)}^{\sigma^{l(f,\tilde \cT) | f}} 
(\prod_{\ell \neq l(f,\tilde \cT) }^{\rightarrow} h_{\ell}^{\sigma^{\ell | f}}) \Big{)} \; .
\eea

Note that if $\cH$ is planar then, for all the leaves of $\tilde \cT$, 
$\prod_{\ell \neq l(f,\tilde \cT) }^{\rightarrow} h_{\ell}^{\sigma^{\ell | f}} $ involves
only lines belonging to $\cT$, hence the relation corresponding to a leaf face implies $h_{l(f,\tilde \cT)}=e$
and, iterating, $h_l=e, \; \forall l \in \cH$.
Using eq. \eqref{eq:routing}, eq. \eqref{eq:ampli} writes 
\bea
A^{ \cG } &=&\frac{(\lambda\bar\lambda)^p}  { [\delta^N(e)]^{p\frac{ (D-2)(D-1)}{2} } }
\int \prod_{\ell\in \cL_{\cG} \setminus  \tilde \cT } dh_{\ell}  
\prod_{ l\in \tilde \cT } d\tilde h_{l } \\
&&\Big{[}  \prod_{f'\notin \cH } 
\delta^N_{f'}( \dots )  \Big{]} \delta^N_{r} ( \dots  ) 
\Big{[} \prod_{f\in \cH , f\neq r} \delta^N_{f} \Big{(} \tilde h_{l(f,\tilde \cT)} \Big{)} \Big{]} 
\nonumber\; , 
\eea
where $\tilde h_{l(f,\tilde \cT)} =  h_{l(f,\tilde \cT)}^{\sigma^{l(f,\tilde \cT) | f}} 
(\prod_{\ell \neq l(f,\tilde \cT) }^{\rightarrow} h_{\ell}^{\sigma^{\ell | f}}) $. Integrating 
$\tilde h_{l(f,\tilde \cT)} $ and bounding the remaining delta functions by $\delta^N(e)$ we obtain
a {\it jacket bound} \cite{colorN} 
\bea\label{eq:jacketbound}
 A^{ \cG } \le  (\lambda\bar\lambda)^p [\delta^N(e)]^{ -p\frac{(D-2)(D-1)}{2} + \cF_{\cG} -\cF_{\cH} +1 } \; ,
\eea 
and $A^{\cG}$ saturates eq. \eqref{eq:jacketbound} if $\cH$ is planar.
The number of lines of $\cG$ is $\cL_{\cG}=(D+1)p$ hence the number of faces of a 
jacket $\cJ$ (with genus $g_{\cJ}$) is $\cF_{\cJ} = (D-1) p + 2-2g_{\cJ}$.
Taking into account that $\cG$ has $\frac{1}{2} D!$ jackets and each face belongs to $(D-1)!$ jackets, 
\bea\label{eq:faces}
 \cF_{\cG} = \frac{D (D-1)}{2} p + D - \frac{2}{(D-1)!} \sum_{\cJ} g_{\cJ} \; ,
\eea
and eq. \eqref{eq:jacketbound} translates into 
\bea\label{eq:boundfin}
A^{\cG} &\le&  (\lambda\bar\lambda)^p [\delta^N(e)]^{   D - 1 - \frac{2}{(D-1)!} \sum_{\cJ} g_{\cJ} + 2 g_{\cH}   }
\crcr
&\le& (\lambda\bar\lambda)^p [\delta^N(e)]^{   D - 1 - \frac{2(D-2)}{D!} \sum_{\cJ} g_{\cJ}  }
\; ,
\eea
where for the last inequality we chose $\cH$ with $g_{\cH} = \inf_{\cJ} g_{\cJ}$. 
We conclude
\begin{itemize}
 \item A graph $\cG$ having {\it at least} a non planar jacket has amplitude bounded by 
$A^{\cG} \le (\lambda\bar\lambda)^p [\delta^N(e)]^{   D - 1 - \frac{2(D-2)}{D!}}$ .
 \item A graph $\cG$ whose {\it all} jackets are planar saturates the bound \eqref{eq:boundfin}, 
  $A^{\cG} = (\lambda\bar\lambda)^p [\delta^N(e)]^{   D - 1 }$, and contributes to the leading order in
  $1/N$. 
\end{itemize}

A first example of a graph whose all jackets are planar is obtained by reconnecting the two lines of color $i$
in the graph $\cG_1$ drawn in figure \ref{fig:melon}. We will denote this graph 
$\cS$. Remark that $\cS$ is dual to two $D$-simplices identified coherently along their $D-1$ 
boundary simplices i.e. the sphere $S^D$. Our result, eq. \eqref{main}, is achieved by the following theorem.
\begin{theorem} \label{lem:sph}
 If all the jackets $\cJ$ of a $D+1$ colored graph $\cG$ are planar then the graph is dual to a sphere $S^D$.
\end{theorem}

\noindent{\bf Proof:}  The theorem obviously 
holds for $D=2$. Eq. \eqref{eq:routing} implies 
$h_l=e, \; \forall l \in \cG$, hence the dual of $\cG$ is simply connected. In $D=3$ one concludes
by the Poincar\'e-Perelman theorem that
$\cG$ is dual to a sphere. But in $D\ge 4$ one should check that {\it all} homotopy groups of 
the dual of $\cG$ coincide with the ones of the sphere, which is cumbersome. 
Let's use instead induction on $D$.

\medskip

{\it Step 1: Contracting a full set of 1-Dipoles.} As all the jackets $\cJ$ are planar, by lemma \ref{lem:planar} all
$\cJ^{\widehat{i}}_{(\rho)}$ are planar, hence (by the induction hypothesis) all $\cB^{\widehat{i}}_{(\rho)}$
are dual to spheres $S^{D-1}$. 

Considering the $D$-bubbles $B^{\widehat{0}}_{(\rho)}$ as effective vertices one can choose 
a ``connectivity tree'' $\cT^0$ of lines of color $0$ (a set of lines connecting all $B^{\widehat{0}}_{(\rho)}$
without forming loops) \cite{colorN}. All the lines in $\cT^0$ are 1-Dipoles and can be contracted, hence
$\cG$ is equivalent with a graph with only one $D$-bubble $\cB^{\widehat{0}}$. 
Any further contractions of 1-Dipoles of colors $1,2,\dots D$ cannot disconnect $\cB^{\widehat{0}}$. 
The genus of the jackets $\cJ$ does not change under 1-Dipole contractions  (the number of vertices 
lines and faces of any jacket decreases by $2$, $D+1$ and $D-1$ respectively).

Iterating for all colors, $\cG\sim\cG'$ with $\cG'$ a graph having exactly one $D$-bubble 
$\cB^{\widehat{i}}, \; \forall i$ and only planar jackets.

\medskip

{\it Step 2: $\cG' = \cS$.} Consider first an arbitrary graph $\cG$. Each of its bubbles
$\cB^{\widehat{i}}_{(\rho)}$ (with $\cN_{ \cB^{\widehat{i}}_{(\rho)} } = 2 p^{ \widehat{i}}_{(\rho)}$)
is a $D$-colored graph hence by eq. \eqref{eq:faces} 
\bea\label{eq:faces2}
 \cF_{ \cB^{\widehat{i}}_{(\rho)} } &=& \frac{(D-1)(D-2)}{2} p^{ \widehat{i}}_{(\rho)} 
+ (D-1) \crcr
&-& \frac{2}{(D-2)!} \sum_{\cJ^{ \widehat{i}}_{(\rho)}  } g_{\cJ^{ \widehat{i}}_{(\rho)}}\; .
\eea
Each vertex of $\cG$ contributes to $D+1$ of its $D$-bubbles ($\sum_{i;\rho} p^{\widehat{i}}_{\rho} = (D+1)p$), 
and each face to $D-1$ of them. Adding eq. \eqref{eq:faces2} yields
\bea
 \cF_{\cG} &=& \frac{(D-2)(D+1)}{2} p \crcr 
    &+& \sum_{i;\rho} \Big{(} 1 - \frac{2}{(D-1)!}
 \sum_{\cJ^{\widehat{i}}_{(\rho)} } g_{ \cJ^{\widehat{i}}_{(\rho)} } \Big{)} \; ,
\eea
which equated with \eqref{eq:faces} writes
\bea\label{eq:genus}
\sum_{\cJ} g_{\cJ} &=& \frac{D!}{2} + \frac{(D-1)!}{2} p \crcr
   &-& \sum_{i;\rho} \Big{(} 
\frac{(D-1)!}{2} -  \sum_{\cJ^{\widehat{i}}_{(\rho)} } g_{ \cJ^{\widehat{i}}_{(\rho)} }
\Big{)} \; .
\eea
For the graph $\cG'$ we have $g_{\cJ}=g_{ \cJ^{\widehat{i}}_{(\rho)}}=0$ and, for all $i$, the sum
over $\rho$ has a single term. It follows that $\cG'$ has exactly two vertices ($p=1$) hence $\cG'=\cS$.
Thus $\cG\sim \cS$ and the dual of $\cG$ is homeomorphic to the sphere $S^D$.

\qed

In $D=3$ we have $\cB^{\widehat{i}}_{(\rho)}= \cJ^{\widehat{i}}_{(\rho)}$ and
eq. \eqref{eq:genus} becomes $\sum_{\cJ} g_{\cJ} = 3 + p -\sum_{i;\rho} \Big{(} 
1 -  g_{ \cB^{\widehat{i}}_{(\rho)} } \Big{)} $. The key to the full topological 
expansion of \cite{colorN} lies in the fact that one can always cancel most planar 
bubbles by 1-Dipole moves. The final graphs obtained by this procedure
(Core Graphs \cite{colorN}) admit a bound in the number of vertices and
index topologies.

\section*{Acknowledgements}

The authors would like to thank Valentin Bonzom for recalling us the subtleties of the Poincar\'e
conjecture in $D\ge 4$.
Research at Perimeter Institute is supported by the Government of Canada through Industry 
Canada and by the Province of Ontario through the Ministry of Research and Innovation.
 Research
by V. Rivasseau is supported by ANR grant LQG-09.

\end{document}